\documentclass[5p,authoryear,twocolumn]{elsarticle}
\usepackage{graphicx,natbib}
\usepackage{amsmath,amssymb,latexsym}
\usepackage{multirow}
\usepackage{natbib}
\usepackage{color}
\newcommand{\Teff}{T_{\rm eff}}

\newcommand{\Leff}{L_{\rm eff}}
\newcommand{\Leq}{L_{\rm eq}}
\newcommand{\Eint}{E_{\rm int}}
\newcommand{\ME}{M_{\oplus}}
\newcommand{\RE}{R_{\oplus}}
\newcommand{\MN}{M_{\rm N}}
\newcommand{\RN}{R_{\rm N}}
\newcommand{\MU}{M_{\rm U}}
\newcommand{\RU}{R_{\rm U}}
\newcommand{\Req}{R_{\rm eq}}
\newcommand{\Rmean}{R_{\rm mean}}
\newcommand{\Zsol}{Z_{\odot}}
\newcommand{\gccm}{g$\,$cm$^{-3}$}
\newcommand{\Tcore}{T_{\rm c}}
\newcommand{\Pcore}{P_{\rm c}}
\newcommand{\Pcenter}{P_{0}}
\newcommand{\Mcore}{M_{\rm c}}
\newcommand{\Ptrans}{P_{\rm 1-2}}
\newcommand{\mcutoff}{m_{\rm cut}}
\newcommand{\Fcond}{F_{\rm cond, ad}}
\newcommand{\Fconv}{F_{\rm conv}}
\newcommand{\fig}{Fig.$\:$}

\newcommand{\sect}{\S$\:$}
\newcommand{\nI}{\mbox{$\lambda$}}

\begin{document}
\begin{frontmatter}
\title{New indication for a dichotomy in the interior structure of Uranus and Neptune 
from the application of modified shape and rotation data\\\vspace{0.1cm}
\footnotesize\textit{Accepted to Planet. Space Sci., June 2012}}
 
\author[URos]{N.~Nettelmann}
\author[UTelA]{R.~Helled}
\author[UCSC]{J.~J.~Fortney}
\author[URos]{R.~Redmer}
\address[URos]
{Institute for Physics, University of Rostock, 18051 Rostock, Germany}
\address[UTelA]
{Department of Geophysics and Planetary Science, Tel-Aviv University Ramt-Aviv, Tel-Aviv 61390, Israel} 
\address[UCSC]
{Department of Astronomy and Astrophysics, University of California, Santa Cruz, CA 95064, USA}

\begin{abstract}
Since the Voyager fly-bys of Uranus and Neptune, improved gravity field data have been derived from 
long-term observations of the planets' satellite motions, and modified shape and solid-body rotation 
periods were suggested. A faster rotation period ($-40$min) for Uranus and a slower rotation period 
($+1$h20) of Neptune compared to the Voyager data were found to minimize the dynamical heights and wind speeds. 
We apply the improved gravity data, the modified shape and rotation data, and the physical LM-R equation of 
state to compute adiabatic three-layer structure models, where rocks are confined to the core, 
and homogeneous thermal evolution models of Uranus and Neptune. We present the full range of structure 
models for both the Voyager and the modified shape and rotation data. 
In contrast to previous studies based solely on the Voyager data or on empirical EOS, we find that 
Uranus and Neptune may differ to an observationally significant level in their atmospheric heavy element mass
fraction $Z_1$ and nondimensional moment of inertia, \nI. 
For Uranus, we find $Z_1\leq 8$\% and $\nI=0.2224(1)$, while for Neptune $Z_1\leq 65$\% and 
$\nI=0.2555(2)$ when applying the modified shape and rotation data, while for the unmodified data we compute 
$Z_1\leq 17$\% and $\nI=0.230(1)$ for Uranus and $Z_1\leq 54$\% and $\nI=0.2410(8)$ for Neptune.
In each of these cases, solar metallicity models $(Z_1=0.015)$ are still possible.
The cooling times obtained for each planet are similar to recent calculations with the Voyager rotation 
periods: Neptune's luminosity can be explained by assuming an adiabatic interior while Uranus cools far too slowly. 
More accurate determinations of these planets' gravity fields, shapes, rotation periods, atmospheric heavy 
element abundances, and intrinsic luminosities are essential for improving our understanding of the internal 
structure and evolution of icy planets.
\end{abstract}

\end{frontmatter}

\section{Introduction}

The outer planets Uranus and Neptune are mysterious in many ways. While their names \emph{'ice giants'} 
suggest a composition of predominantly volatiles in ice phases such as water, methane, and ammonia ice, 
interior models instead predict a warm interior devoid of solid ices \citep{Chau+11,Redmer+11}. 
Structure models of Uranus and Neptune are generally in agreement in predicting a small rock core, 
a deep interior of more than 70\% heavy elements, and a significantly less enriched
outer envelope with a transition at about 70\% of the radius for both planets \citep{Hubbard+95,FN10,Helled+11}. 
However, they also agree in failing to explain Uranus' measured intrinsic luminosity \citep{Podolak+91}.
Uranus' low luminosity is a riddle; even more so as the corresponding models for
Neptune nowadays can reproduce its measured luminosity \citep{Fortney+11} indicating that the
ice giants are not so similar to each other as previously thought \citep{Podolak+95}. 

Contrary to the efforts that have been made to measure the atmospheric composition, the Voyager~2 radio 
occultation data and ground-based observational microwave data have actually raised more questions 
than they were intended to solve. In both planetary atmospheres for instance, D/H appears enriched
over the protosolar values \citep{Bergh86,Bergh90} suggesting particle transport between the atmosphere 
and an ice-rich deep interior \citep{Hubbard+95}, while the inferred helium abundance is consistent 
with the protosolar value and thus would suggest the opposite (i.e., inefficient particle transport)
if hydrogen and helium occur in the deep interior where they might undergo phase separation 
\citep{Hubbard+95,Podolak+95}. Moreover, for the interior of Uranus we cannot rule out a primary 
composition of silicates dissolved into hydrogen--helium envelopes as that might sufficiently 
accelerate Uranus' cooling time \citep{HMacF80}. We just do not know how similar Uranus and Neptune 
really are  in terms of composition and structure \citep{Podolak+00}. 

In addition, Uranus and Neptune have complex multi-polar magnetic fields, and appear to have
stronger atmospheric winds than Jupiter and Saturn, if the Voyager rotation periods are applied.
Besides the natural desire to understand the giant planets in our solar system, 
learning about the 'ice giants' is important for the classification of extrasolar planets with 
similar masses and sizes, as many ones are observed \citep{Borucki+11}, although not yet on 
similar orbital distances \citep{Kane11}.

Information on Uranus and Neptune interiors are typically derived from theoretical models which are 
designed to fit the observed physical data of the planets, such as their gravitational fields, 
masses, internal rotation, and radii. The physical data available for Uranus and Neptune 
are rather limited. 
In particular, the low-order gravitational harmonics $J_2$ and $J_4$  of Neptune have significant 
error bars, Neptune's equatorial 1-bar radius is actually not measured, and the shapes (flattening) of both 
Uranus and Neptune are not well known. Voyager 2 provided only one occultation radius at 
the 1 bar pressure level for each planet \citep[see][for details]{Helled+10}. Although stellar 
occultations do provide information on the planetary oblateness, the shape is inferred for upper 
atmosphere (microbar pressure levels) and it is unclear whether this shape is consistent with the 
shape at the 1 bar pressure level. \citet{Helled+10} have shown that minimization of wind velocities 
or dynamic heights of the 1 bar isosurfaces \citep{AS07,Helled+09} constrained by Voyager 2 occultation 
radii and gravitational coefficients of the planets, leads to \emph{modified solid-body rotation periods} of 
16h 34m for Uranus and 17h 27m for Neptune, which is 40~min shorter than the Voyager rotation period for Uranus, 
and 1h20 longer for Neptune. 
\citet{Helled+10} also state that both planets may have different rotation periods than the 
ones derived by the minimization method, and in addition, could be rotating differentially on cylinders.
Non-solid body rotation can lead to a change in the calculated gravitational moments \citep{Hubbard+91}. 
Our results are valid as long as differential rotation in Uranus and Neptune is "shallow", i.e., 
the region of differential rotation consists of a negligible mass, which is a preferred solution 
for Uranus and Neptune (Y. Kaspi, priv.~comm.).
However, the suggested solid-body rotation periods better match the measured radii of the planets and 
result in more moderate atmosphere dynamics with wind velocities of $\sim$ 150 m s$^{-1}$ for both planets.
Based on these suggested rotation rates and occultation radii, \emph{modified
shapes} of the planets were derived. 
Since the shapes of Uranus and Neptune are not well constrained, and in addition, their rotation profiles 
must not be that of a solid-body, the modified solid-body rotation periods and shapes can be used to 
demonstrate the sensitivity of the interior models to the uncertainties in these physical properties.

To obtain pressure-density relations for the interior of Uranus and Neptune, three different methods have been invoked. \citet{Podolak+95} apply physical equations of state of hydrogen, helium, the ices H$_2$O, H$_2$S, CH$_4$, and NH$_3$, 
and rocks, assuming sharp transitions between a gaseous outer envelope, an inner icy envelope, and a rock
core. As this assumption by itself restricts the possible internal density distributions, \citet{Marley+95} and
\citet{Podolak+00} created random density distributions which, when reproducing the observed gravity field,
could then be evaluated according to pressure-density relations of likely materials. Another possibility is 
to use an analytic function with sufficiently many free coefficients to adjust to the given constraints 
\citep{Helled+10}. The resulting pressure-density relation can be considered an \emph{empirical} EOS.
The advantage of the two latter methods is that they can allow for, and one can constrain the locations of, continuous (rather than sharp) density gradients. 
On the other hand, the resulting random or fitted density distributions can be unrealistic
in a sense of not representing any composition of real materials. We here use the physical EOS LM-REOS 
for H, He, and H$_2$O \citep{N+08} for the envelope material and the rock EOS by \citet{HM89} for the core. 
This combination was also applied to the Uranus and Neptune models in \citet{FN10} (hereafter FN10).

In this paper, we use the suggested modified periods and equatorial radii  
and the physical LM-R EOS to compute new three-layer interior and evolution models 
of Uranus and Neptune. In \sect\ref{sec:m} we describe our method of model construction
and the observational data used. In \sect\ref{sec:r} we compare the resulting models
obtained with the modified data to those with the Voyager data. The results are discussed and
summarized in \sect\ref{sec:dc}.

\section{Method of Structure and Evolution Modeling}\label{sec:m}

Our Uranus and Neptune models consist of three layers; a rocky core surrounded by two adiabatic 
and homogeneous envelopes of hydrogen, helium, and water. 
The interior model uses the following physical data: the planetary total mass $M_p$, 
the equatorial radius $\Req$, the temperature at the 1 bar pressure level $T_1$ which 
determines the internal adiabat, the rotation rate $\omega=2\pi/P$ with $P$ being the solid-body 
rotation period, and the measured gravitational coefficients $J_2$, $J_4$. The physical parameters 
used in the models are listed in Table~\ref{tab:obs}.
For simplicity, it is assumed that both Uranus and Neptune are adiabatic and convective, 
although Uranus may not have an adiabatic temperature gradient (Podolak et al., 1991). 
The resulting model parameters of interest here are the mass fraction 
of heavy elements in the inner envelope, $Z_2$, and in the outer envelope, $Z_1$, where we
restrict $Z_1$ to $\geq 1\:\Zsol$, $\Zsol=0.015$ being the solar metallicity \citep{Lodders03},
the core mass, $\Mcore$, and the possible transition pressure $\Ptrans$ between the envelopes.
The resulting values $Z_1$, $Z_2$, and $\Mcore$ are obtained by an iterative scheme that fits 
$J_2$, $J_4$ and the condition of mass conservation $m=0$ at the center $r=0$. 
A constant He:H mass ratio of 0.275:0.725 is assumed in the envelopes.
The method used to calculate interior models is the same as in (\citealt{N+08}, FN10) and the 
thermal evolution procedure the same as in FN10.
The progress over that work is mainly the application of improved observational data.
Recent work by Jacobson and collaborators provided new estimates 
for the gravitational harmonics of both Uranus and Neptune using Earth-based 
astrometry and observations acquired with the Voyager spacecraft. Their more accurate determinations
included more astrometric observations over the years, a change in reference frame to the International
Celestial Reference Frame, a modification in the numerically integrated Proteus orbit (for Neptune), 
and an improved data processing procedure (see \citet{Jacobson09} for details).
As in \citet{Helled+09,Helled+10} we here use the \citet{Jacobson07,Jacobson09} data\footnote{see also
the NASA/JPL website\\ {\tt http://ssd.jpl.nasa.gov/?gravity\underline{\ }fields\underline{\ }op}} 
for the gravitational moments $J_2$ and $J_4$. These are provided at reference equatorial 
radii as given in Table~1.
The gravitational moments at the equatorial 1-bar radius to be fitted by our models are then obtained 
from the measured values (primed quantities) by scaling according to the relation 
$\Req^{2n}J_{2n} = \Req'^{2n}J_{2n}'$. Figure~1 shows these improved data in comparison with the
former Voyager gravity data, and the relative observational uncertainties for all four outer
planets.

\hspace*{1cm}\\
{\bf[Table 1]}\\
{\bf[Figure 1]}

For the thermal evolution we make the standard assumption that the planet cools down over time 
by the release of gravitational and internal energy remnant from their formation, where the heat from
the interior is transported by convection along an adiabatic temperature gradient. 
Thus our assumption of different heavy element mass fractions in the two envelopes, 
($Z_1\neq Z_2$) implies that one compositional gradient exists that inhibits 
convection across the layer boundary, but not the heat flow.

With the \emph{cooling time $\tau$} we denote the time the planet needs to cool down from an arbitrarily
hot initial state to its current effective temperature $\Teff$. To calculate $\tau$, we pick a 
certain structure model (meaning a set of stationary parameters 
\{$Z_1$, $Z_2$, $m_{\rm 1-2}$, $\Mcore$\}, where $m_{\rm 1-2}=m(\Ptrans)$), and calculate $\sim 60$ 
internal profiles with stepwise increased $T_1$ values, that represent the planet at 
earlier times. We then integrate the cooling equation $\Leff-\Leq = -d\Eint\:/\:dt$
over time, where $\Leff$ is the measured infrared luminosity, $\Leq$ the insolation, 
and $d\Eint$ the lost intrinsic energy during the time interval $dt$. To close the relations, 
we use the model atmosphere approximation by \citet{Guillot+95}, and the Stefan-Boltzmann 
law to relate $T_1$ to $\Leff$. 
While the underlying grid for non-irradiated, solar metallicity model atmospheres by 
\citet{Graboske+75} may appear inappropriate for application to Uranus and Neptune despite its 
former application (\citealp{HMacF80}, FN10), the induced uncertainty in the resulting cooling 
times of Uranus and Neptune compared to those based on a more self-consistent grid for Uranus and Neptune 
has been found to be $\sim 0.2$ Gyr only \citep{Fortney+11}. Both grids neglect the presence of 
clouds and hazes which are thought to be present in the tropospheres \citep{Gautier+95}. 
We note that clouds are just at the beginning of being incorporated into climate models for the Earth
\citep{Dessler10}. 
Therefore, the full uncertainty on the cooling time from the application of the model atmosphere grid is
not known but future work on cloudy skies is not expected to significantly alter the main findings 
of this work as we are searching for differences between Uranus and Neptune.
This also justifies the neglection of the time-dependence of the insolation, of the
rotation rate due to angular momentum conservation, and of the energy of rotation. Their effects on the
cooling time have been shown to nearly compensate each other in case of Jupiter \citep{N+12}.

\section{Results}\label{sec:r}

\subsection{Models with the Voyager shape and rotation data}\label{ssec:r_un1}

Figure \ref{fg:unZZ1} shows our results for the envelope metallicities of three-layer Uranus and 
Neptune models with the Voyager shape and rotation data. 
There are no solutions with $Z_1\geq \Zsol$ outside the respective boxes, as either the core
 mass becomes zero, or $|J_4|$ becomes too low or too high.

Models with the same transition pressure show a
linear dependence between $Z_1$ and $Z_2$, because to first order,  the resulting decrease in the  
outer envelope's mass ($dM_1$) is proportional to the imposed reduction $-dZ_1$. Mass conservation 
then requires $dM_2+d\Mcore=-dM_1$, where $dM_2\sim dZ_2$. Consequently, $dZ_2\sim -dZ_1$. 
If mass is shifted downward ($dM_1<0$), $|J_4|$ must decrease. Thus a lower limit of $Z_1$ at 
given $\Ptrans$ arises from the condition that $|J_4|$ must not become too small. Since $\Mcore$ 
decreases with increasing $Z_2$ in standard three-layer models as known for Jupiter \citep{N+12}, 
at very high $Z_2$ values ($\gtrsim 0.95$) the core mass decreases to zero before the lower 
observational limit of $|J_4|$ is reached. Thus the Neptune models along the upper right boundary 
in \fig\ref{fg:unZZ1} are in fact two-layer models with no core.
The internal  adiabat then reaches temperatures up to 6600~K and pressures up to 8~Mbar, 
while the lowest core-mantle boundary temperature, $\Tcore$ (pressure, $\Pcore$) is 5200~K (5.5~Mbar), 
found for the model with the biggest rock core of mass $\Mcore=3.7\:\ME$.

\begin{figure}
\includegraphics[width=0.8\textwidth]{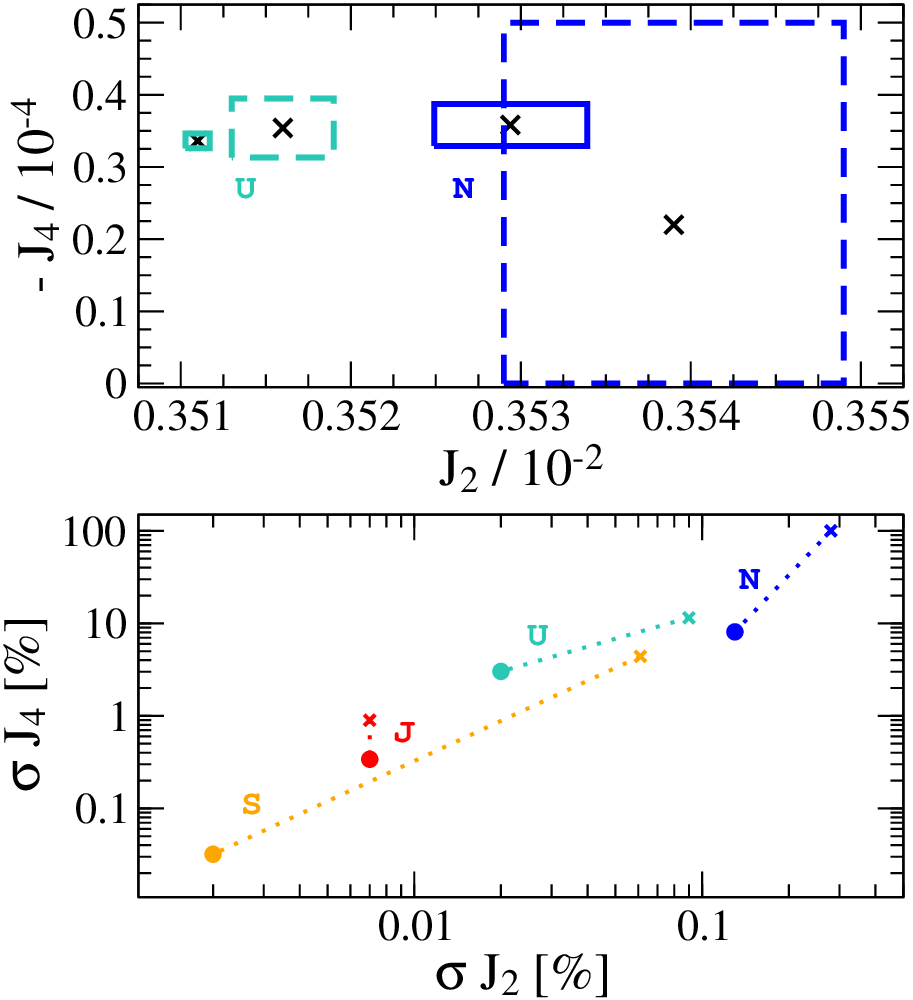}
\caption{\label{fg:obsJ2J4}(Color online)}
\begin{minipage}{0.8\textwidth}
\vspace*{0.5cm}
\emph{Upper panel:} Observed gravitational moments $J_2$ and $J_4$ of Uranus (U, \emph{cyan}) and Neptune 
(N, \emph{blue}) with $\mathbf 1\sigma$ error bars according to the Voyager and pre-Voyager data 
(\emph{dashed}) and the Jacobson data (\emph{solid}). Crosses indicate the mean values. 
\emph{Lower panel:} relative observational $1\sigma$ uncertainties 
in $J_2$ and $J_4$ from the Voyager and Pioneer missions (\emph{crosses}) and improved values (\emph{circles}); 
\emph{J, red}: Jupiter, improved Galileo data from \citet{Jacobson03}; \emph{S, orange:} Saturn, improved Cassini data 
from \citet{AS07}; \emph{U, cyan:} Uranus; \emph{N, blue:} Neptune.
\end{minipage}
\end{figure}

\begin{figure}
\centering
\includegraphics[width=0.8\textwidth]{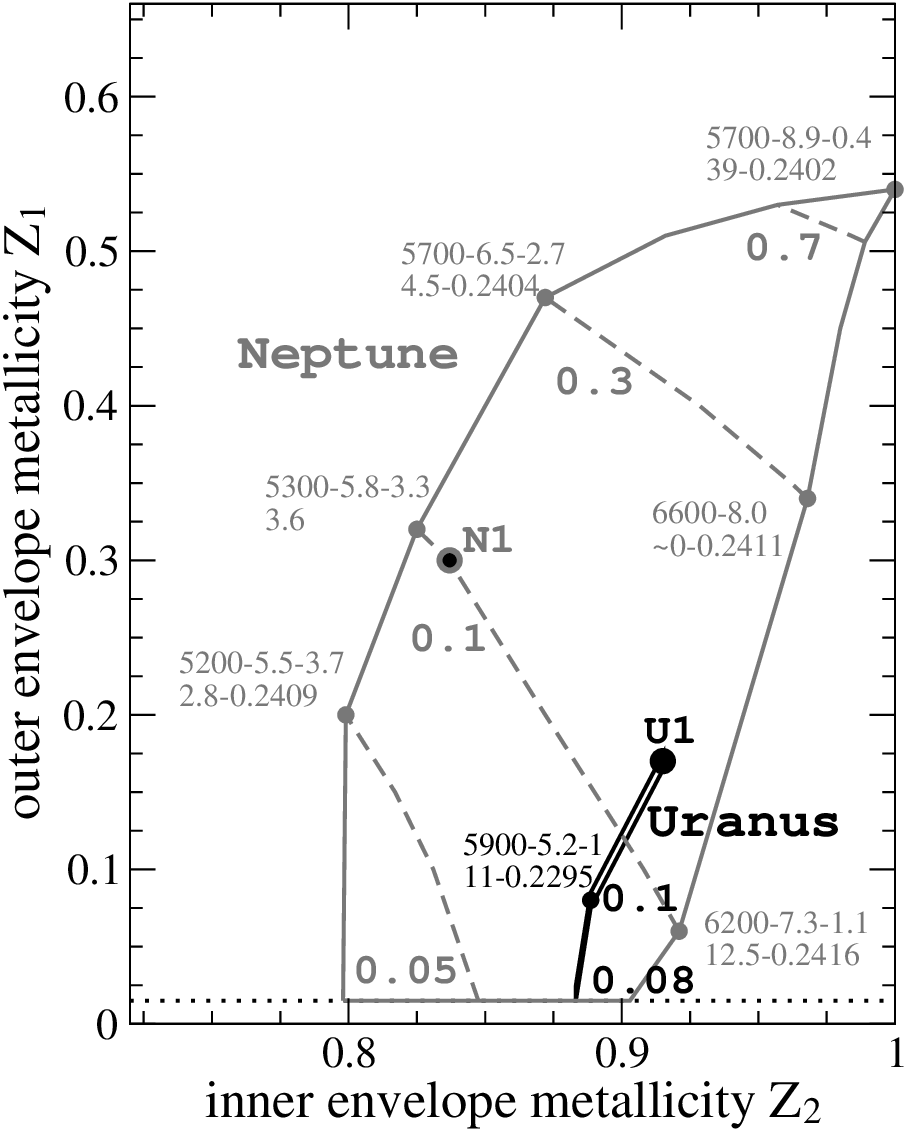}
\caption{\label{fg:unZZ1}(Color online)} 
\begin{minipage}{0.8\textwidth}
\vspace*{0.5cm}
Heavy element mass fraction in the outer envelope ($Z_1$) and inner envelope ($Z_2$) of Uranus 
models (\emph{black}) and Neptune models (\emph{grey}) as labeled with the Voyager shape and rotation data. 
The \emph{solid} lines frame the full set of solutions for each planet. Dashed lines within the 
box of Neptune models indicate solutions of same transition pressure in [Mbar] as labeled. 
Numbers at selected models (\emph{filled circles}) give $\Tcore$~[K], $\Pcore$~[Mbar], 
$\Mcore\:[\ME]$, the ice-to-tock ratio I:R, and $\nI$.
The \emph{dotted} line is a guide to the eye for the solar metallicity $Z_{\odot}=0.015$.
\end{minipage}
\end{figure}

The described behavior is the same as found previously for Uranus and Neptune (see figure 3 in FN10). 
However, due to the large observational error bars of $J_2$ and $J_4$ (\fig\ref{fg:obsJ2J4})
allowed for in FN10, the Neptune models were arbitrarily forced to have $Z_1\leq 0.40$ and 
$\Ptrans\geq 0.15$~Mbar in accordance with the results for Uranus. It is the much -reduced error
bars of $J_2$ and $J_4$ that here allow us to present the full range of models. The biggest 
difference to the set of Neptune models in FN10 is the existence of models with pure water envelopes. 
Those occur for high outer envelope metallicities $Z_1>0.5$. Then, $\Ptrans$ needs to be put 
deep inside the planet $(P\sim 1$~Mbar) to ensure that $|J_4|$ does not get too large.

Few Uranus models can be found for the improved error bars of $J_2$ and $J_4$ (\fig\ref{fg:obsJ2J4}). 
They have a well defined $Z_2=0.9\pm 0.02$, a tightly constraint position of the layer boundary at 
$0.12\pm 0.04$~Mbar corresponding to $r=0.79 \pm 0.02\: \RU$ and $m=0.93\pm 0.02\: \MU$, 
a rather low outer envelope metallicity $1\:\Zsol\leq Z_1\leq 12\:\Zsol$, a rock core of at most 
$1\:\ME$, and central temperatures $\Tcore\sim 6000$~K. An example is model U1 (Table~2).
In $Z_1$--$Z_2$ space, the Uranus models are within the range of the Neptune models.

\hspace*{1cm}\\
{\bf[Figure 2, top left]}\\
{\bf[Figure 3, top right]}\\

\subsection{Models with the modified shape and rotation data}\label{ssec:r_un2}

\begin{figure}
\centering
\includegraphics[width=0.8\textwidth]{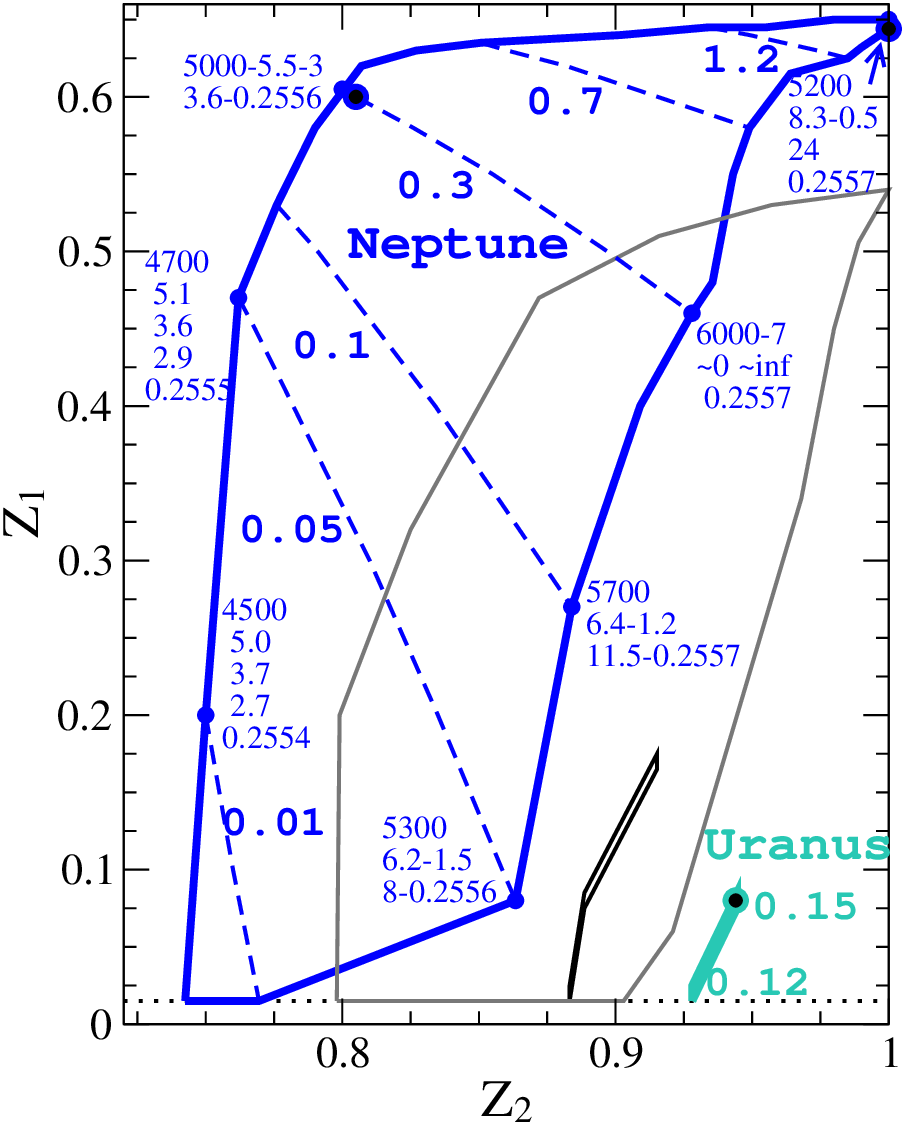}
\caption{\label{fg:unZZ2}(Color online)}
\begin{minipage}{0.8\textwidth}
\vspace*{0.5cm}
Same as \fig\ref{fg:unZZ1} but using the modified shape and rotation data for Uranus (\emph{cyan})
and Neptune (\emph{blue}). Models U2, N2a, and N2b (Table~2) are highlighted by 
\emph{black-filled, big circles}. 
The boxes of \fig\ref{fg:unZZ1} are also shown to facilitate the comparison.
\end{minipage}
\end{figure}

The explanations to \fig\ref{fg:unZZ1} in \sect\ref{ssec:r_un1} apply also to \fig\ref{fg:unZZ2}. 
We here describe the differences that arise from using the modified shape and rotation data.

Most obvious, the sets of solutions shift into opposite directions: the Uranus models move to the
lower right corner of \fig\ref{fg:unZZ2} where $Z_2$ gets higher and $Z_1$ smaller, increasing the
density difference between the envelopes. In contrast, the Neptune models stretch into the upper
left corner where $Z_1$ is above 60\% and the density difference to the deep envelope less pronounced.
As a result, the sets of solutions become disjunct.
The Uranus solutions are tightly constrained to have a low outer envelope metallicity ($<0.1$), 
a high inner envelope metallicity ($>0.9$), a transition far out (at $>0.9\:\MU$), 
and a small core ($< 1\ME$). An example is model U2, see Table~2. 
Similar Uranus and Neptune solutions are still possible, for instance with $Z_1\sim 0.1$ and $Z_2\sim 0.88$ 
in both planets. However, this would require an envelope transition in Neptune at $\sim 0.05$~Mbar, 
relatively far out at $0.88\:\RN$ in the neutral, molecular fluid part, which is not a preferred 
solution for Neptune.
The new finding here is that two kinds of qualitatively different Neptune models are possible that both
differ in the atmospheric heavy element abundance to an observationally significant level from the Uranus models. 
The first kind of Neptune models are characterized by a rather large core ($\Mcore\sim 3\ME$) and
a modest heavy element difference between the envelopes (changing from from 0.6 to 0.8). 
Alternatively, small core models are possible with pure water envelopes and a transition deep inside at 
$\sim 0.6\:\RN$ and $\sim 0.6\:\MN$. Models N2a and N2b are respective examples (Table~2).

These changes can easily be explained by the new rotation rates.
A slower rotation (Neptune) means a lower centrifugal force. Matter
is then less strongly pushed to the outer region. If the $J_2$, $J_4$ to be fitted
remain about the same, a higher metallicity in the outer part of the planet is required. 
Along the adiabat, a higher metallicity leads to lower internal temperatures. Therefore, 
Neptune models with high outer envelope metallicity can become
rather cold, with core-mantle boundary temperatures around 5000~K only.

\hspace*{1cm}\\
{\bf[Table 2]}

\paragraph{Example density profiles}

\begin{figure}
\centering
\includegraphics[width=0.8\textwidth]{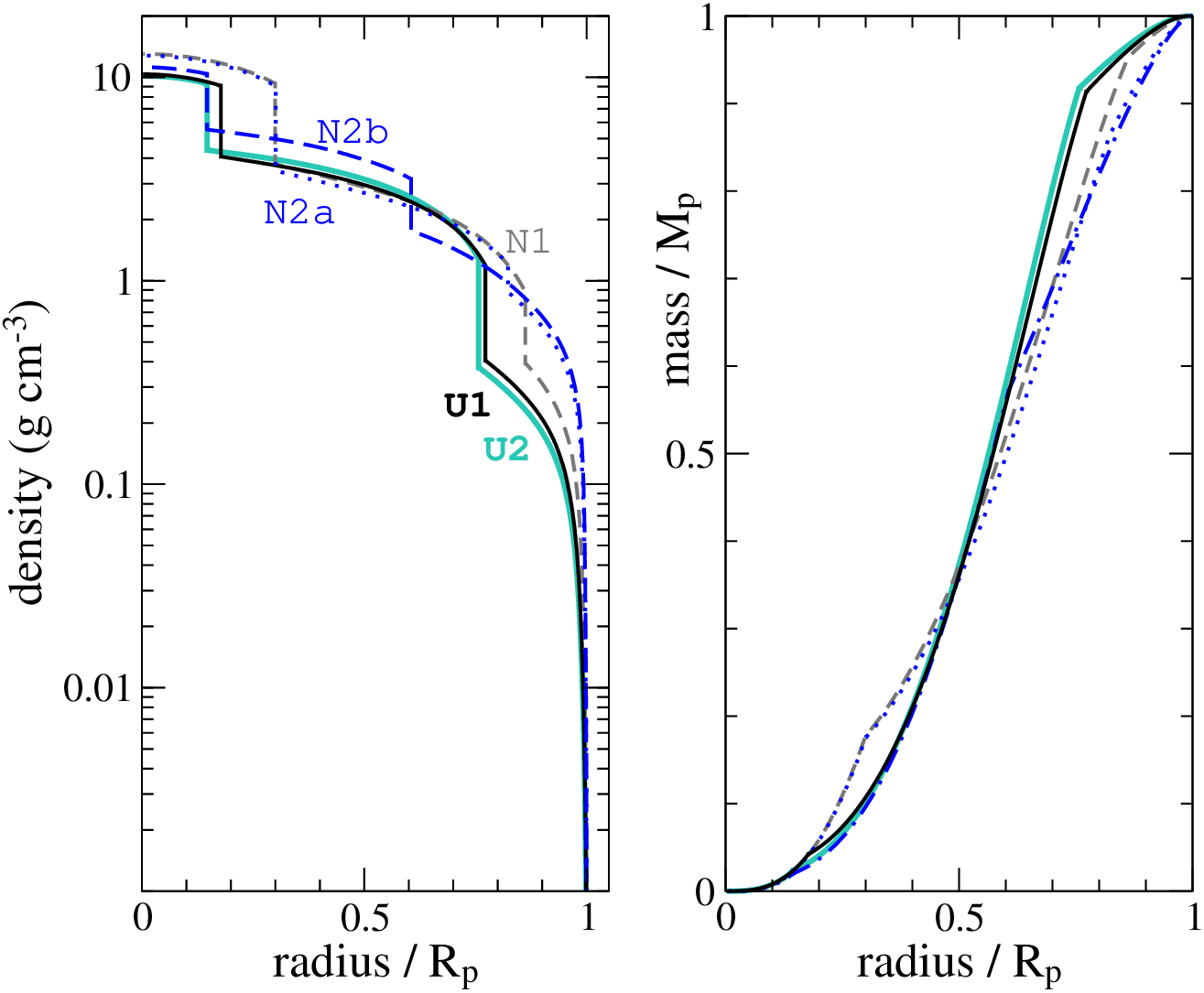}
\caption{\label{fg:profilesR}(Color online)} 
\begin{minipage}{0.8\textwidth}
\vspace*{0.5cm}
Internal density (\emph{left panel}) and mass (\emph{right panel}) profiles over the 
normalized planetary radius of the models U1 (\emph{solid, black}), U2 (\emph{thick solid, cyan}), 
N1 (\emph{short-dashed, grey}), N2a (\emph{dotted, blue}), and N2b (\emph{long-dashed, blue}). 
See Table~2 for details.
\end{minipage}
\end{figure}

\begin{figure}
\centering
\includegraphics[width=0.8\textwidth]{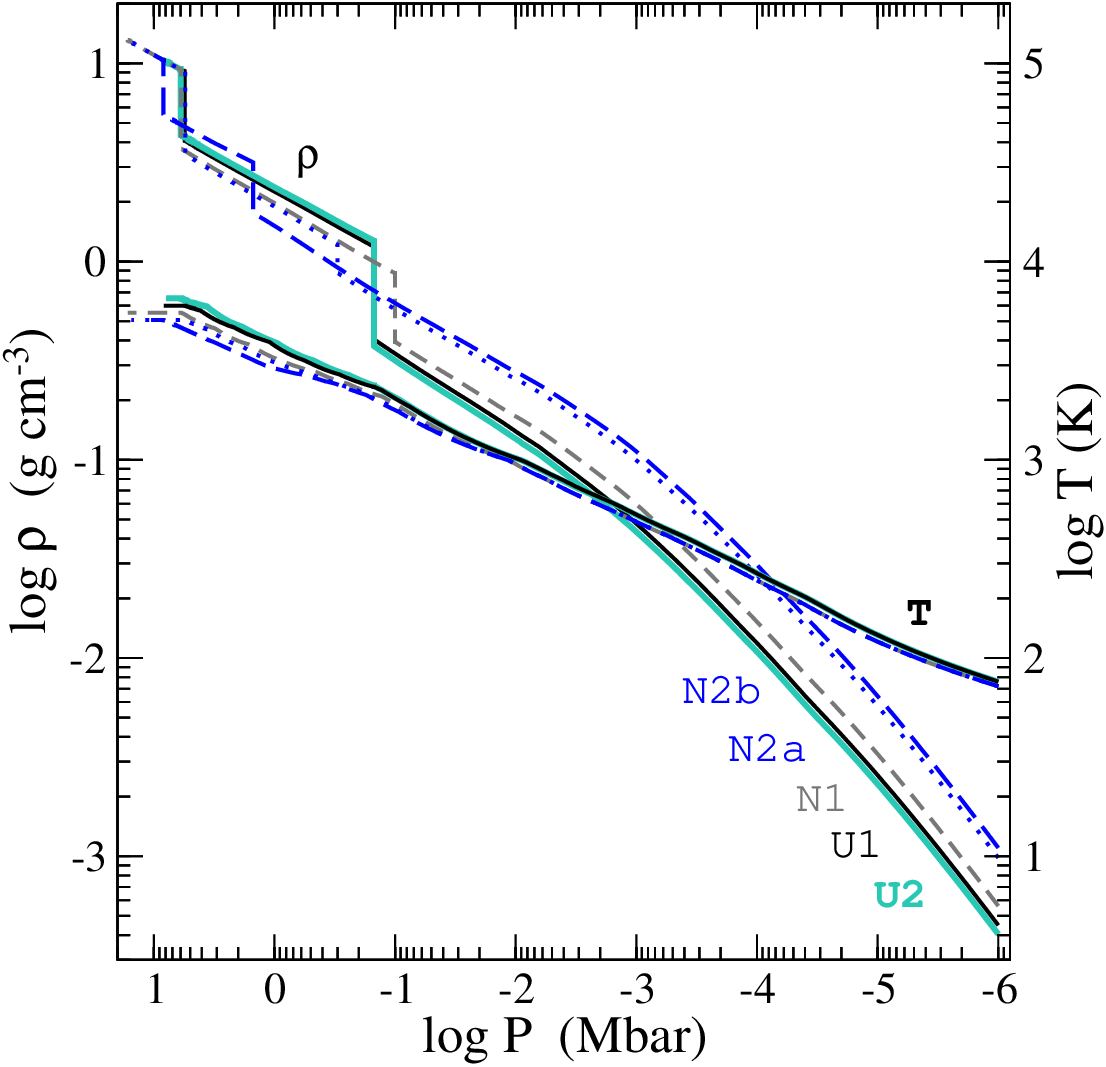}
\caption{\label{fg:profilesLogP}(Color online)} 
\begin{minipage}{0.8\textwidth}
\vspace*{0.5cm}
Internal density and temperature profiles over pressure.
See \fig\ref{fg:profilesR} for description of line-styles and labels.
\end{minipage}
\end{figure}

The input and resulting parameters of representative models are given in Table~\ref{tab2} 
and their density and mass profiles shown in \fig\ref{fg:profilesR}.
The Uranus models clearly stand out by their big jump in density at $\sim 75\%$ of the planet's
radius, where the density falls down to 30\% of the inner envelope boundary value. With the end of the
inner envelope, 90\% of the total mass is already accumulated. For Neptune, a model can be found (N2b)
where the density difference is only 20\% and thus more similar to the transition in standard 
three-layer Jupiter models (10\%, \citealp{N+12}) than to Uranus. For Neptune model N2a, the transition 
occurs deep inside at 60\% of the radius, where 57\% of the total mass is accumulated, which would be 
more similar to standard three-layer models of Saturn \citep{N09} than to any other outer planet.
For both planets, a typical mean density in the inner envelope is 3 \gccm. This (arbitrary) 
density threshold occurs at about $0.5\:R_p$ (\fig\ref{fg:profilesR}, left panel) and $0.4\:M_p$ 
(\fig\ref{fg:profilesR}, right panel), 
and corresponds to pressures of 2--3~Mbars and temperatures of 4000--5000~K (\fig\ref{fg:profilesLogP}).

\hspace*{1cm}\\
\noindent
{\bf[Figure 4]}\\
{\bf[Figure 5]}\\

Machine-readable tabulated interior profiles of mass, pressure, radius, temperature, and density 
of the models U1, U2, N1, and N2b are provided as supplemental online-material to this article. 
Table \ref{tab:SOM} show an example of such a table.

\hspace*{1cm}\\
{\bf[Table 3]}

\paragraph{Moment of Inertia}

The above described dichotomy in the internal structures, which mainly refers to the outer $50\%$ of 
the planet's radius, maps onto different nondimensional moments of inertia\footnote{
$\nI\:=I/(M_p\,\Rmean^2)$, where $I$ is the dimensional axial moment of inertia.}, \nI. 
For Uranus and Neptune, respectively, we calculate values of $\nI=0.2224(1)$ and $\nI=0.2555(2)$
when using the modified shape and rotation data, while in the unmodified case the respective values 
are $\nI=0.230(1)$ and $\nI=0.2410(8)$, where the number in parenthesis gives the uncertainty in the 
last digit, see also Figs.~\ref{fg:unZZ1} and \ref{fg:unZZ2}.


\subsection{Thermal evolution}\label{ssec:evol}

\begin{figure}
\centering
\includegraphics[width=0.8\textwidth]{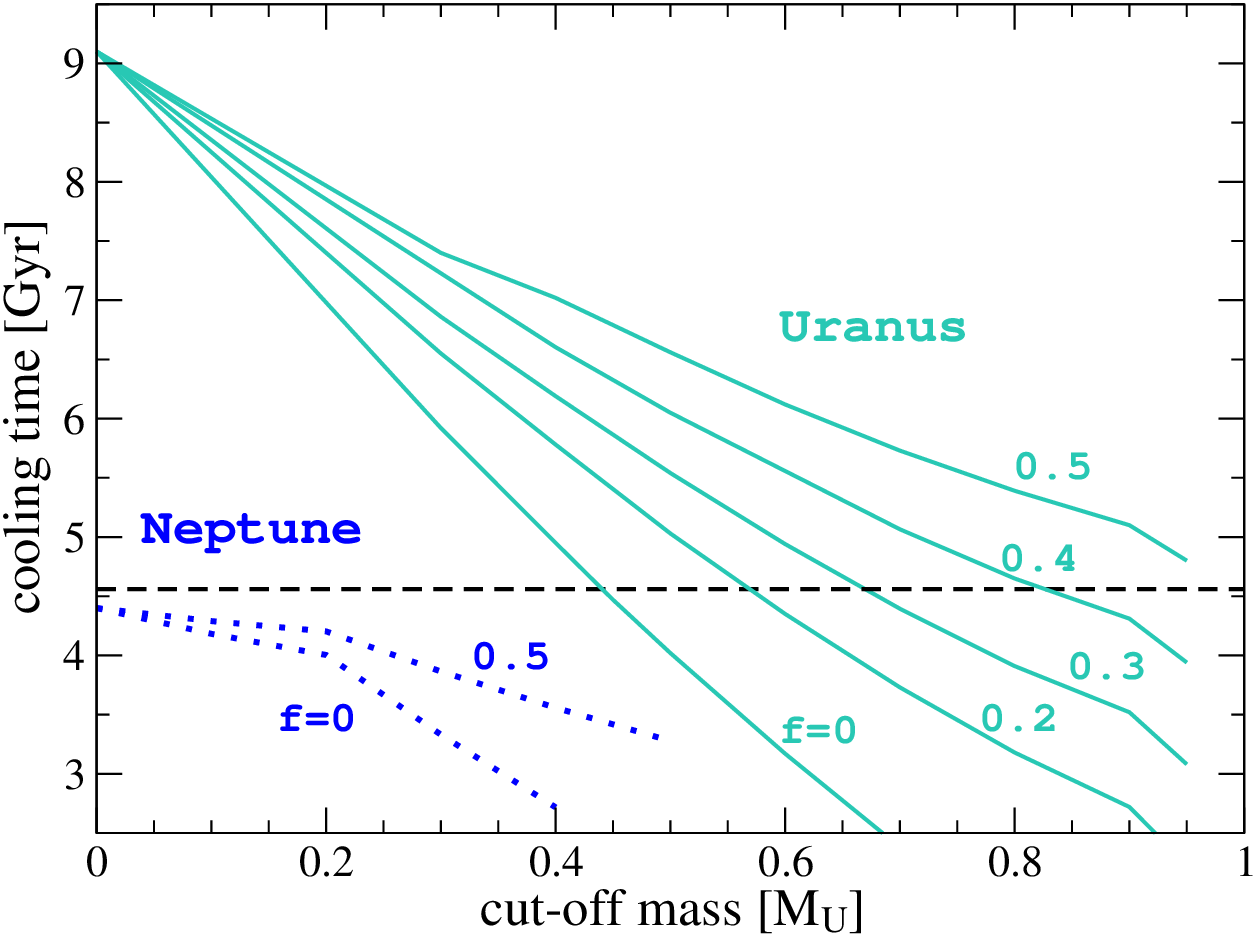}
\caption{\label{fg:evolU}(Color online)} 
\begin{minipage}{0.8\textwidth}
\vspace*{0.5cm}
Thermal evolution of Uranus (model U2, \emph{solid, cyan}) assuming that the heat 
flux from the mass interior to the cut-off mass ($x$-axis) is limited to a fraction $f$ of the 
convective energy flux. For $f=0$--0.4 and a cut-off mass of 0.45--0.85, a cooling time in agreement 
with the age of the solar system (\emph{horizontal dashed line}) can be found.
For comparison, we also show the same calculations for Neptune (model N2a, \emph{dotted, blue}).
\end{minipage}
\end{figure}

The cooling times for the representative models are given in Table~2. The modified 
rotation and shapes have essentially no effect on the cooling time. This is to be expected as
the cooling time mainly depends on the specific heat $c_v$ of the bulk material, while the composition
(and hence $c_v$) is little influenced by the shape and rotation data.
Interestingly, Neptune appears to be the outer planet with the cooling time that best matches the
age of the solar system, $\tau_{\odot}=4.56$~Gyr. Under the assumption of adiabatic, homogeneous cooling, 
Saturn's theoretical cooling time is systematically too short by 2-2.5~Gyrs \citep{Guillot+95,ForHubb03}, while 
Jupiter's is often found to be a little too long \citep{Saumon+92,Guillot+95,Fortney+11}. We here recover the well-known 
far too long cooling time of Uranus \citep{Fortney+11}, indicating that some part of the interior is not 
in a state of efficient energy transport through vigorous convection \citep{Podolak+91,Hubbard+95}. 
To estimate the size of the stable internal region in Uranus, we re-calculate the cooling time for the 
same quasi-adiabatic three-layer models as before but exclude a central mass, the cut-off mass $\mcutoff$, 
from efficient cooling. The efficiency factor $f$ quantifies the heat escape from the stable interior 
beneath $\mcutoff$ in comparison with the convective heat flux, so that $f=0$ implies zero heat flux
and $f=1$, as well as $\mcutoff=0$, is the uninhibited case. In \fig\ref{fg:evolU} we present 
the cooling time of Uranus in dependence on $\mcutoff$ and $f$.

\hspace*{1cm}\\
{\bf[Figure 6]}\\

The larger $\mcutoff$, i.e.~the smaller the convective outer region, the shorter the cooling time. 
For a minimum cut-off mass of $0.45\:\MU$, $\tau=\tau_{\odot}$ can be achieved. The larger the heat
flux from the stable interior, the longer the cooling time, and thus the larger the required cut-off mass.
A maximum value for $f$ is obtained when $\mcutoff$ approaches $\MU$ to realize $\tau=\tau_{\odot}$.
While this maximum is $f=0.5$, we consider small values $f<0.1$ more realistic as the heat transport 
in a stable interior is predicted to be a tiny fraction of the convective heat flux. For Uranus, a lower limit 
for $f$ can be the ratio $\Fcond/\Fconv\approx 0.02$ , where $\Fcond$ is the conductive
heat flux along an adiabatic gradient, and $\Fconv$ the heat flux if Uranus would have cooled down to
its present luminosity by large-scale convection \citep{Podolak+95}.
Thus we conclude from figure \ref{fg:evolU} an ending of the stable interior in Uranus at 0.45$-$0.5~$\MU$. 
For comparison, we have applied the same procedure to Neptune (model N2a). Neptune's cooling time 
appears weakly sensitive to the innermost 20\% of Neptune's mass.

\section{Discussion and Conclusions}\label{sec:dc}

\subsection{Bulk composition}

We had to make several simplifications to be able to calculate Uranus and Neptune
models with state-of-the art methods. One such simplification is the representation of  heavy elements 
in the envelopes by water and the confinement of rocks to the core. In real Uranus and Neptune, 
silicates may also occur in the envelopes. It is clear that our assumption of having no rocks 
in the envelopes leads to an overestimation of the envelope metallicities and 
the resulting ice to rock ratio (I:R). In addition, the smaller the core mass, the larger the 
I:R ratio. Models with I:R $\gg 2.7$, the solar system value, may potentially invalidate this simplification
and point to the presence of silicates in the envelopes. Example models are the Neptune models
with pure water envelopes (N2b) and all our Uranus models. Interestingly, some Neptune models, e.g.~N1 
and N2a, have a rather large core ($\sim 3\ME$) with a reasonable overall I:R ratio of 1.5 times the 
solar value. On the other hand, our representation of ices by a pure water EOS likely overstimates the 
density of the true mixture of ices, and thus somewhat underestimates the mass fraction of ices that would be 
composed of a mixture of H$_2$O, CH$_4$, and NH$_3$. Unfortunately, high-quality EOS of light ices at 
pressures higher than 2~Mbar for planetary modeling are not yet available.

Under the assumptions and simplifications of this work, the bulk mass
of heavy elements is $12.5\:\ME$ for Uranus and 14--$14.5\:\ME$ for the selected Neptune models.
This is in good agreement with the empirical EOS based models by \citet{Helled+11}, who can explain 
the polynomial density distributions of their Uranus (Neptune) models by $\sim 11$--$13\:\ME$ 
($\sim 13$--$15\:\ME$) of heavy elements.

\subsection{Implications from a stable deep interior}

If some part of the interior, for Uranus possibly 0.45--$0.5\:\MU$ (\sect\ref{ssec:evol}), 
is stable to convection so that heat cannot escape efficiently from the region below, 
then the super-adiabaticity of the temperature gradient there can be non-negligible \citep{LC12}.
A warmer deep interior will require a lower particle number density in order to conserve the pressure gradient.
Otherwise, the induced higher warm-dense-matter pressure would cause a larger planet radius. 
For a given composition in the inner envelope (e.g. the $Z_2$ value of the adiabatic case), one might 
think of a lower particle number density to imply a lower envelope mean density (compared to the 
adiabatic case), resulting into a larger rock core mass to ensure mass conservation. 
However, the mean density in the inner envelope is roughly constrained by the measured $J_2$ value,
see \fig\ref{fg:profilesR}.
Therefore, the metallicity in the deep interior cannot have the same $Z_2$ value as in the adiabatic case. Indeed, 
if the stable region is caused by a compositional gradient \citep{Hubbard+95}, the deep interior would have 
an average metallicity $Z_3>Z_2$, which may lead to a smaller rock core. Therefore, predictions on 
a change of the rock core mass are not possible without more sophisticated models that take into 
account both a compositional gradient and super-adiabaticity. At the current level of our models, 
the net effect of a stable interior would be the introduction of a third, deep envelope with $Z_3>Z_2$.
In case of a significant super-adiabaticity in Uranus, this $Z_3$ might raise to 100\%. 
Moreover, a H/He-free deep interior might then be possible even with a material density larger than that
of water, i.e.~with an ice-rock mixture. Thus, a strongly superadiabatic deep interior (below about $0.5\:\MU$) 
may qualitatively allow for I:R ratios closer to the solar and at the same time for a H/He-free deep
interior, resembling a traditional protoplanetary core of a few $\ME$ that can accrete its gaseous envelope
within a few million years \citep{HoriIkoma10}.

A stable deep (i.e., far below the outer/inner envelope boundary) interior offers an 
explanation for the measured atmospheric helium and deuterium abundances. Given that D/H can be 
enriched in the cold ices of the protosolar nebula, the observed enhanced atmospheric D:H ratio could 
result from upward-transport of deuterium from an ice-rich interior \citep{Gautier+95} that is 
located between the inner/outer envelope boundary and the stable deep interior, where the magnetic field
is believed to be generated, see \sect\ref{ssec:dc_lumi}. 
Assuming that hydrogen and helium \emph{exist} in the deep interior at Mbar pressures where hydrogen is metallic, 
\citet{Hubbard+95} suggest that helium could phase separate and rain down leading to a helium-depleted 
atmosphere. However, if the particle exchange with the upper regions is suppressed in the deep interior, 
the rained out helium will not be replenished from the reservoir above so that the atmospheric 
He:H ratio remains protosolar as observed. 
However, we caution against taking the observed D and He abundances for clear indications of a stable deep interior, 
since helium phase separation from hydrogen must not necessarily occur in a metallic environment.
\citet{Lorenzen+09,Lorenzen+11} calculated the H-He immiscibility regions in dependence on temperature, 
pressure, and helium concentration. 
In metallic hydrogen, demixing requires a sufficiently high helium concentration to be energetically 
preferred. For instance, at 2 Mbar our Uranus and Neptune models have a temperature of 3800--4500 K. 
The phase diagram of \citet{Lorenzen+11} predicts demixing under these conditions only if the He 
concentration is above $\sim 1\%$. Using 
\begin{equation}
	\frac{N_{\rm He}}{N} = \left(1+\frac{m_{\rm He}}{m_{\rm H}}\,\frac{X}{Y} 
	+ 3\frac{m_{\rm He}}{m_{\rm H_2O}}\frac{Z}{Y}\right)^{-1}\:,
\end{equation}
where the $m_{i}$ denote the molar masses of H, He, and water, we calculate helium particle
concentrations $N_{\rm He}/N$ of 1.7--3\% for $Z=0.9$--0.95. This is only slightly above the demixing condition in pure
H-He mixtures. As the demixing behavior of helium in more complex mixtures such as H-He-H$_2$O is unknown,
we cannot rule out miscibility of He the deep interior of Uranus and Neptune, be it stable (Uranus case)
or unstable (Neptune case) as an explanation for the observed abundances. We encourage future work on the
miscibility behavior of planetary mixtures, such as started by, e.g., \citet{Chau+11,WM12a,WM12b}.

\subsection{Luminosity}\label{ssec:dc_lumi}

The modified shape and rotation data do not affect the cooling behavior. 
With Uranus, this is to be expected, since former calculations with various 
different structure models as allowed by the large error bars in $J_2$ and $J_4$ produced an 
uncertainty of about 1 Gyr in the evolution only, smaller than the uncertainty induced by the 
observational error bar of $\Teff$ \citep{Fortney+11}. 
With Neptune, this is of some surprise, since the new models are $\sim 1000\:$K colder in the deep 
interior than former models used for cooling curve calculations (FN10). However, the cooling curve 
depends on changes of temperature with time. These are similar right after the rapid cooling at young ages.

As the heat stored in the interior after formation depends on the specific heat of the bulk
material, where $c_v$ is smaller for rocks than for ices, future work should include the
admixture of rocks into the envelopes and investigate the maximum possible shortening of Uranus' 
cooling time within the approach of homogeneous, adiabatic evolution. 

Magnetic field models of Uranus and Neptune require that 60--70\% of the region interior 
to the ionic water layer, corresponding to 0.42--0.56 $R_p$ \citep{Redmer+11}, is stable to 
convection \citep{StanBlox06}.  Using our interior models, this corresponds to 0.25--0.5 
of the planet's mass. Indeed we can reproduce Uranus' measured luminosity if we assume that 
the heat flux from within 0.45--0.5$\:M_p$ is a negligible fraction of the heat flux in case of a 
convective, adiabatic interior. This consistency was already noted by \citet{Podolak+91}.
However, this mass or radius level is not supported by any of our Uranus structure models. 
Even if a density gradient occurs continuously rather than as a sharp transition, our structure 
models for Uranus predict a location farther out at $0.9\MU$.
Also for Neptune, a stable interior up to 0.25--0.5\:$M_p$ is not supported
by our structure models, where the deepest strong density gradient can occur at 0.57$\MN$.
Neptune's luminosity is best explained by the absence of a stable region.
Future work should aim at finding consistent solutions for the structure, thermal evolution, and
and magnetic field generation of Uranus and Neptune. At the current stage, the envelope separation
(density gradient) of our structure models does not seem to be directly related to the luminosity, 
nor to the magnetic field generation.

\subsection{Why a dichotomy?}

Beside the low luminosity, Uranus differs from Neptune in having a high obliquity, dense narrow rings 
and its five largest satellites on regular orbits, while  Neptune has a more Saturn-like obliquity and 
extended dusk disk with diffuse rings, and two major satellites (Triton and Nereide) on irregular orbits.
These observed properties point to different formation histories and as a consequence, also to different
internal structures. As discussed by \citet{Stevenson86}, the stochastic process 
of impacts of various sizes and obliquities may have caused the differences we see today.
A composition gradient in a heavy-element-rich giant planet could be a remnant of the formation process 
\citep{Podolak+91} that has been disturbed in Neptune by a last big direct impact but not so in Uranus 
(an oblique impact).
We are not the first authors to suggest a dichotomy between the internal structures of Uranus and Neptune. 
But our structure models are the first ones to confirm this pre-Voyager hypothesis of their different 
structures also on the basis of computed moment of inertia values, with an absolute difference of 
$d\nI\sim 0.03\:(\sim 0.01)$ for the models with the (un-)modified shape and rotation data.

\subsection{A Vote for improved observational data}

New measurements of Uranus and Neptune's physical parameters could improve our knowledge of their bulk
compositions and internal structures considerably. For instance, the outer envelope 
metallicity is rather sensitive to the assumed solid-body rotation period. 
While the modified periods used in this paper are perhaps not the correct ones for Uranus and Neptune, 
since differential rotation is possible and the winds may not be in a state of minimum energy, 
the presented models clearly indicate the need for more data. 
Recently, \citet{Karkoschka11} analyzed a collection of atmospheric circulation data
compiled from Voyager~2 and HST observations of Neptune's atmosphere over a time-span of 20 years
and found a rotation period close to the Voyager value, based on the stability of the motion of
high-altitude clouds.
Better determination of the rotation periods, shapes, Uranus' luminosity, Neptune's gravity 
field, and of the envelope metallicities below the water cloud level are indispensable for 
a better understanding of the interior, evolution, and formation of the icy planets in the 
outer solar system. We encourage the application of various discovery methods in the future, 
and in particular, to fly space missions to Uranus and Neptune. 
  
\subsection{A dichotomy in the atmosphere compositions?}

One of the indications for a dichotomy in the interior structure is the 
possible difference in the outer envelope metallicities $Z_1$, which are at most 0.08 ($\leq0.18$) 
for Uranus but up to 0.65 ($\leq0.54$) for Neptune,  using the (un-)modified shape 
and rotation data. In case of vertical mixing up into the troposphere,  one would expect to see a 
signature from different outer envelope metallicities in the lower atmosphere. Over the past 
60 years, several attempts of deriving the atmosphere abundances from measured spectra have 
been undertaken, see, e.g., the reviews of \citet{Fegley+91,Gautier+95,GuiGau07}. 
However, that turned out to be quite challenging, mostly because of the numerous absorption lines
of CH$_4$ and H$_2$, uncertainties in the line shapes and absorption coefficients, non-equilibrium chemistry,
and the dependences of the derived abundances among each other. Today, 
carbon is the heavy element for which the interpretation of the available data gives the least
ambiguous picture. A value of C:H=30--60$\times$ solar for both planets has finally emerged \citep{Gautier+95}. 
Assuming C:H$=30\times$ solar and similar enrichments also of N and O, \citet{Hubbard+95} 
estimate an outer envelope ice mass fraction of 0.08. While this is in good agreement with 
the possible $Z_1$ values of our Uranus and Neptune models, it does not support our high-$Z_1$ 
Neptune models as required for qualitatively different interiors. On the other hand, both CO and HCN have been detected in the atmosphere of Neptune but not of Uranus \citep{Gautier+95}, indicating
different processes at work in their atmospheres. To replace assumptions by real data, 
we encourage deep entry probe missions to Uranus and Neptune.

\subsection{Past and present structure models}

As shown in figures \ref{fg:obsJ2J4}--\ref{fg:unZZ2}, more definite conclusions about the dissimilarity 
of the internal structures are prevented by the current observational uncertainty in the 
gravitational moments. This is at odds with the struggles of earlier modelers \citep{Podolak+95,Marley+95}
to find Uranus and Neptune models at all that matched the less accurately known gravity data in the past. 
Moreover, despite the tighter present constraints, our models seem to encompass many of the previously 
found models. 

In particular, \citet{HM89} developed Uranus models with physical EOS, smooth transitions between 
an outer H-He rich envelope, an inner ice-rich envelope, and a rock core, and showed that models 
with 5--15\% H-He in the inner envelope could give best agreement with the imposed constraints. 
Such an H-He abundance is also seen in our Uranus models. 

When the Voyager gravity and rotation rate data for Neptune became available, \citet{Podolak+95} 
applied the three-layer approach and physical EOS to compute Uranus and Neptune models. Their Uranus models
required some H-He to be present in the deep interior, too, whereas for Neptune this was found to be optional
(non-conventional models). They could also find a different Neptune model with an inner envelope of 
pure water and transition deeper inside at $\sim 0.7\RN$, and a highly enriched outer envelope, indicating
a possible difference in the interior structures of Uranus and Neptune. 
Our Neptune models in the upper right corners of Figs.~\ref{fg:unZZ1},\ref{fg:unZZ2} are similar to 
that latter one, while their non-conventional models 
are within the bulk of our Uranus and Neptune models.

\citet{Marley+95} again allowed for smooth density gradients and generated models with random density 
distributions. Nevertheless, they found that a rather sharp transition from a low-density outer envelope 
to a high-density inner envelope is necessary to fit the \emph{Voyager} values of $J_2$ and $J_4$, 
where the pressure-density relation in the inner envelope could be that of an ice layer. Because of the 
admittance of smooth density gradients, they found for the first time that the transition could occur 
as deep as between $\sim$ 60--65\% of the planet's radius. 
Our model N2b shows the same properties, although within the three-layer approach. 

Finally, \citet{Helled+11} were the first to use the accurate post-Voyager gravity data of 
\citet{Jacobson07,Jacobson09}. Their polynomial density-profiles could well represent a metallicity 
gradient that rises from about solar metallicity at the surface to about 85\% in the center, 
with slightly higher values preferred for Neptune but no indication of different internal structures. 
The found metallicities are within those of our models.

According to the present work, the application of the modified shape and rotation data gives a more 
pronounced indication of different structure than seen so far.

\subsection{Conclusions}

We present the full sets of three-layer interior models (with H/He/water envelopes and rocks
confined to the core) of Uranus and Neptune for different solid-body rotation periods and flattenings, 
using the improved gravity data by \citet{Jacobson07,Jacobson09}, and physical equations of state. 
We find that the resulting bulk composition is insensitive to the current level of uncertainty in
the input data (observational constraints and equations of state) as our results are in good agreement 
with previous calculations (e.g.~\citealt{Podolak+95, Marley+95, Helled+11}).
However, our models with the modified rotation periods and shapes suggest that Uranus 
and Neptune could be quite different. Uranus would have an outer envelope with a few times
the solar metallicity which transitions to a heavily enriched ($\sim 90\%$ by mass heavy elements)
inner envelope at $0.9\:\MU$, giving a rather low moment of inertia of $\sim 0.222$. 
In Neptune, this transition can occur deeper inside at $0.6\:\MN$ and be accompanied by a more 
moderate increase in metallicity, leading to a less centrally condensed planet with $\nI\sim 0.255$.
While the observed magnetic fields of Uranus and Neptune are similar and can be reproduced by a rather 
narrow range of dynamo models, dissimilar interiors are required to explain the measured luminosities. 
We have presented a new indication for different internal structures based on the
application of modified shape and rotation data. However, the density gradient in our models 
appears to be generally farther out than required by evolution and magnetic field models.

The authors thank the two referees for providing fruitful comments that led us to consider 
the results in a wider context. NN acknowledges support from the DFG RE 881/11-1.

\bibliographystyle{model2-names}
\bibliography{msUNinterior-refs}

\begin{thebibliography}{47}
\expandafter\ifx\csname natexlab\endcsname\relax\def\natexlab#1{#1}\fi
\expandafter\ifx\csname url\endcsname\relax
  \def\url#1{\texttt{#1}}\fi
\expandafter\ifx\csname urlprefix\endcsname\relax\def\urlprefix{URL }\fi
\providecommand{\eprint}[2][]{\url{#2}}
\providecommand{\bibinfo}[2]{#2}
\ifx\xfnm\relax \def\xfnm[#1]{\unskip,\space#1}\fi
\bibitem[{Anderson and Schubert(2007)}]{AS07}
\bibinfo{author}{Anderson, J.}, \bibinfo{author}{Schubert, G.},
  \bibinfo{year}{2007}.
\newblock \bibinfo{title}{{Saturn's Gravitational Field, Internal Rotation, and
  Interior Structure}}.
\newblock \bibinfo{journal}{Science} \bibinfo{volume}{317},
  \bibinfo{pages}{1384--1387}.
\bibitem[{de~Bergh et~al.(1986)de~Bergh, Lutz, Owen, Brault and
  Chauville}]{Bergh86}
\bibinfo{author}{de~Bergh, C.}, \bibinfo{author}{Lutz, B.L.},
  \bibinfo{author}{Owen, T.}, \bibinfo{author}{Brault, J.},
  \bibinfo{author}{Chauville, J.}, \bibinfo{year}{1986}.
\newblock \bibinfo{title}{Monoteuterated methane in the outer solar system. ii.
  its detection on uranus at 1.6 $\mu$m}.
\newblock \bibinfo{journal}{ApJ} \bibinfo{volume}{311},
  \bibinfo{pages}{501--510}.
\bibitem[{de~Bergh et~al.(1990)de~Bergh, Lutz, Owen and Maillard}]{Bergh90}
\bibinfo{author}{de~Bergh, C.}, \bibinfo{author}{Lutz, B.L.},
  \bibinfo{author}{Owen, T.}, \bibinfo{author}{Maillard, J.P.},
  \bibinfo{year}{1990}.
\newblock \bibinfo{title}{Monoteuterated methane in the outer solar system. iv.
  its detection and abundance on neptune}.
\newblock \bibinfo{journal}{ApJ} \bibinfo{volume}{355},
  \bibinfo{pages}{661--666}.
\bibitem[{Borucki et~al.(2011)Borucki, Koch, Basri, Batalha, Brown, Bryson,
  Caldwell, Christensen-Dalsgaard, Cochran and {et al.}}]{Borucki+11}
\bibinfo{author}{Borucki, W.J.}, \bibinfo{author}{Koch, D.},
  \bibinfo{author}{Basri, G.}, \bibinfo{author}{Batalha, N.},
  \bibinfo{author}{Brown, T.}, \bibinfo{author}{Bryson, S.T.},
  \bibinfo{author}{Caldwell, D.}, \bibinfo{author}{Christensen-Dalsgaard, J.},
  \bibinfo{author}{Cochran, W.D.}, \bibinfo{author}{{et al.}},
  \bibinfo{year}{2011}.
\newblock \bibinfo{title}{{Characteristics of Planetary Candidates Observed by
  Kepler. II. Analysis of the First Four Months of Data}}.
\newblock \bibinfo{journal}{ApJ} \bibinfo{volume}{736}, \bibinfo{pages}{A19}.
\bibitem[{Brozovic and Jacobson(2009)}]{BJ09}
\bibinfo{author}{Brozovic, M.}, \bibinfo{author}{Jacobson, R.A.},
  \bibinfo{year}{2009}.
\newblock \bibinfo{title}{{The Orbits of the Outer Uranian Satellites}}.
\newblock \bibinfo{journal}{Astronom. J.} \bibinfo{volume}{137},
  \bibinfo{pages}{3834}.
\bibitem[{Chau et~al.(2011)Chau, Hamel and Nellis}]{Chau+11}
\bibinfo{author}{Chau, R.}, \bibinfo{author}{Hamel, S.},
  \bibinfo{author}{Nellis, W.J.}, \bibinfo{year}{2011}.
\newblock \bibinfo{title}{{Chemical processes in the deep interior of Uranus}}.
\newblock \bibinfo{journal}{Nature Communications} \bibinfo{volume}{2},
  \bibinfo{pages}{A203}.
\bibitem[{Dessler(2010)}]{Dessler10}
\bibinfo{author}{Dessler, A.E.}, \bibinfo{year}{2010}.
\newblock \bibinfo{title}{A determination of the cloud feedback from climate
  variations of the last decade.}
\newblock \bibinfo{journal}{Science} \bibinfo{volume}{330},
  \bibinfo{pages}{1523}.
\bibitem[{Fegley et~al.(1991)Fegley, Gautier, Owen and Prinn}]{Fegley+91}
\bibinfo{author}{Fegley, B.}, \bibinfo{author}{Gautier, D.},
  \bibinfo{author}{Owen, T.}, \bibinfo{author}{Prinn, R.G.},
  \bibinfo{year}{1991}.
\newblock \bibinfo{title}{Spectroscopy and chemistry of the atmosphere of
  uranus}, in: \bibinfo{editor}{Bergstrahl, J.T.}, \bibinfo{editor}{Miner,
  E.D.}, \bibinfo{editor}{Matthews, M.S.} (Eds.), \bibinfo{booktitle}{Uranus}.
  \bibinfo{publisher}{University of Arizona, Tucson}, p. \bibinfo{pages}{147}.
\bibitem[{Fortney and Hubbard(2003)}]{ForHubb03}
\bibinfo{author}{Fortney, J.J.}, \bibinfo{author}{Hubbard, W.B.},
  \bibinfo{year}{2003}.
\newblock \bibinfo{title}{{Phase Separation in Giant Planets: Inhomogeneous
  evolution of Saturn}}.
\newblock \bibinfo{journal}{Icarus} \bibinfo{volume}{164},
  \bibinfo{pages}{228}.
\bibitem[{Fortney et~al.(2011)Fortney, Ikoma, Nettelmann, Guillot and
  Marley}]{Fortney+11}
\bibinfo{author}{Fortney, J.J.}, \bibinfo{author}{Ikoma, M.},
  \bibinfo{author}{Nettelmann, N.}, \bibinfo{author}{Guillot, T.},
  \bibinfo{author}{Marley, M.S.}, \bibinfo{year}{2011}.
\newblock \bibinfo{title}{{Self-consistent Model Atmospheres and the Cooling of
  the Solar System Giant Planets}}.
\newblock \bibinfo{journal}{ApJ} \bibinfo{volume}{729}, \bibinfo{pages}{32}.
\bibitem[{Fortney and Nettelmann(2010)}]{FN10}
\bibinfo{author}{Fortney, J.J.}, \bibinfo{author}{Nettelmann, N.},
  \bibinfo{year}{2010}.
\newblock \bibinfo{title}{{The Interior Structure, Composition, and Evolution
  of Giant Planets}}.
\newblock \bibinfo{journal}{Space Sci.~Rev.} \bibinfo{volume}{152},
  \bibinfo{pages}{423--447}.
\bibitem[{Gautier et~al.(1995)Gautier, Conrath, Owen, "De~Pater" and
  Atreya}]{Gautier+95}
\bibinfo{author}{Gautier, D.}, \bibinfo{author}{Conrath, B.J.},
  \bibinfo{author}{Owen, T.}, \bibinfo{author}{"De~Pater", I.},
  \bibinfo{author}{Atreya, S.K.}, \bibinfo{year}{1995}.
\newblock \bibinfo{title}{The troposphere of neptune}, in:
  \bibinfo{editor}{Cruishank} (Ed.), \bibinfo{booktitle}{Neptune and Triton}.
  \bibinfo{publisher}{University of Arizona, Tucson}, pp.
  \bibinfo{pages}{547--611}.
\bibitem[{Graboske et~al.(1975)Graboske, Plness, Pollack and
  Grossman}]{Graboske+75}
\bibinfo{author}{Graboske, H.C.}, \bibinfo{author}{Plness, R.J.},
  \bibinfo{author}{Pollack, J.B.}, \bibinfo{author}{Grossman, A.S.},
  \bibinfo{year}{1975}.
\newblock \bibinfo{title}{The structure and evolution of jupiter: the fluid
  contraction stage}.
\newblock \bibinfo{journal}{ApJ} \bibinfo{volume}{199}, \bibinfo{pages}{265}.
\bibitem[{Guillot et~al.(1995)Guillot, Chabrier, Gautier and
  Morel}]{Guillot+95}
\bibinfo{author}{Guillot, T.}, \bibinfo{author}{Chabrier, G.},
  \bibinfo{author}{Gautier, D.}, \bibinfo{author}{Morel, P.},
  \bibinfo{year}{1995}.
\newblock \bibinfo{title}{{Effect of radiative transport on the evolution of
  Jupiter and Saturn}}.
\newblock \bibinfo{journal}{ApJ} \bibinfo{volume}{450}, \bibinfo{pages}{463}.
\bibitem[{Guillot and Gautier(2007)}]{GuiGau07}
\bibinfo{author}{Guillot, T.}, \bibinfo{author}{Gautier, D.},
  \bibinfo{year}{2007}.
\newblock \bibinfo{title}{The giant planets}, in: \bibinfo{editor}{Schubert,
  G.}, \bibinfo{editor}{Spohn, T.} (Eds.), \bibinfo{booktitle}{Treatise of
  Geophysics, vol.~10, Planets and Moons}. \bibinfo{publisher}{Amsterdam:
  Elsevier}, p. \bibinfo{pages}{439 (arXiv:0912:2019)}.
\bibitem[{Helled et~al.(2011)Helled, Anderson, Podolak and
  Schubert}]{Helled+11}
\bibinfo{author}{Helled, R.}, \bibinfo{author}{Anderson, J.},
  \bibinfo{author}{Podolak, M.}, \bibinfo{author}{Schubert, G.},
  \bibinfo{year}{2011}.
\newblock \bibinfo{title}{{Interior models of Uranus and Neptune}}.
\newblock \bibinfo{journal}{Astrophys. J.} \bibinfo{volume}{726},
  \bibinfo{pages}{A15}.
\bibitem[{Helled et~al.(2010)Helled, Anderson and Schubert}]{Helled+10}
\bibinfo{author}{Helled, R.}, \bibinfo{author}{Anderson, J.D.},
  \bibinfo{author}{Schubert, G.}, \bibinfo{year}{2010}.
\newblock \bibinfo{title}{{Uranus and Neptune: shape and rotation}}.
\newblock \bibinfo{journal}{Icarus} \bibinfo{volume}{210},
  \bibinfo{pages}{446}.
\bibitem[{Helled et~al.(2009)Helled, Schubert and Anderson}]{Helled+09}
\bibinfo{author}{Helled, R.}, \bibinfo{author}{Schubert, G.},
  \bibinfo{author}{Anderson, J.D.}, \bibinfo{year}{2009}.
\newblock \bibinfo{title}{{Jupiter and Saturn Rotation Periods}}.
\newblock \bibinfo{journal}{Planet. Space Sci.} \bibinfo{volume}{57},
  \bibinfo{pages}{1467--1473}.
\bibitem[{Hori and Ikoma(2010)}]{HoriIkoma10}
\bibinfo{author}{Hori, Y.}, \bibinfo{author}{Ikoma, M.}, \bibinfo{year}{2010}.
\newblock \bibinfo{title}{Critical core masses for gas giant formation with
  grain-free envelopes}.
\newblock \bibinfo{journal}{ApJ} \bibinfo{volume}{714}, \bibinfo{pages}{1343}.
\bibitem[{Hubbard and MacFarlane(1980)}]{HMacF80}
\bibinfo{author}{Hubbard, W.B.}, \bibinfo{author}{MacFarlane, J.J.},
  \bibinfo{year}{1980}.
\newblock \bibinfo{title}{{Structure and Evolution of Uranus and Neptune}}.
\newblock \bibinfo{journal}{J. Geophys. Res.} \bibinfo{volume}{88},
  \bibinfo{pages}{225}.
\bibitem[{Hubbard and Marley(1989)}]{HM89}
\bibinfo{author}{Hubbard, W.B.}, \bibinfo{author}{Marley, M.S.},
  \bibinfo{year}{1989}.
\newblock \bibinfo{title}{{Optimized Jupiter, Saturn, and Uranus interior
  models}}.
\newblock \bibinfo{journal}{Icarus} \bibinfo{volume}{78}, \bibinfo{pages}{102}.
\bibitem[{Hubbard et~al.(1991)Hubbard, Nellis, Mitchell, Holmes, McCandless and
  Limaye}]{Hubbard+91}
\bibinfo{author}{Hubbard, W.B.}, \bibinfo{author}{Nellis, W.J.},
  \bibinfo{author}{Mitchell, A.C.}, \bibinfo{author}{Holmes, N.C.},
  \bibinfo{author}{McCandless, P.C.}, \bibinfo{author}{Limaye, S.S.},
  \bibinfo{year}{1991}.
\newblock \bibinfo{title}{Interior structure of neptune - comparison with
  uranus}.
\newblock \bibinfo{journal}{Science} \bibinfo{volume}{253},
  \bibinfo{pages}{648--651}.
\bibitem[{Hubbard et~al.(1995)Hubbard, Podolak and Stevenson}]{Hubbard+95}
\bibinfo{author}{Hubbard, W.B.}, \bibinfo{author}{Podolak, M.},
  \bibinfo{author}{Stevenson, D.J.}, \bibinfo{year}{1995}.
\newblock \bibinfo{title}{{The Interior of Neptune}}, in:
  \bibinfo{editor}{Cruishank} (Ed.), \bibinfo{booktitle}{Neptune and Triton}.
  \bibinfo{publisher}{University of Arizona, Tucson}, p. \bibinfo{pages}{109}.
\bibitem[{Jacobson(2003)}]{Jacobson03}
\bibinfo{author}{Jacobson, R.A.}, \bibinfo{year}{2003}.
\newblock \bibinfo{howpublished}{{JUP230 orbit solution}}.
\bibitem[{Jacobson(2007)}]{Jacobson07}
\bibinfo{author}{Jacobson, R.A.}, \bibinfo{year}{2007}.
\newblock \bibinfo{title}{{The Gravity Field of the Uranian System and the
  Orbits of the Uranian Satellites and Rings}}.
\newblock \bibinfo{journal}{BAAS} \bibinfo{volume}{39}, \bibinfo{pages}{453}.
\bibitem[{Jacobson(2009)}]{Jacobson09}
\bibinfo{author}{Jacobson, R.A.}, \bibinfo{year}{2009}.
\newblock \bibinfo{title}{{The Orbits of the Neptunian Satellites and the
  Orientation of the Pole of Neptune}}.
\newblock \bibinfo{journal}{Astronomical J.} \bibinfo{volume}{137},
  \bibinfo{pages}{4322}.
\bibitem[{Kane(2011)}]{Kane11}
\bibinfo{author}{Kane, S.R.}, \bibinfo{year}{2011}.
\newblock \bibinfo{title}{{Detecting the Signatures of Uranus and Neptune}}.
\newblock \bibinfo{journal}{Icarus} \bibinfo{volume}{214},
  \bibinfo{pages}{327--333}.
\bibitem[{Karkoschka(2011)}]{Karkoschka11}
\bibinfo{author}{Karkoschka, E.}, \bibinfo{year}{2011}.
\newblock \bibinfo{title}{Neptune's rotational period suggested by the
  extraordinary stability of two features}.
\newblock \bibinfo{journal}{Icarus} \bibinfo{volume}{215},
  \bibinfo{pages}{439--448}.
\bibitem[{Leconte and Chabrier(2012)}]{LC12}
\bibinfo{author}{Leconte, J.}, \bibinfo{author}{Chabrier, G.},
  \bibinfo{year}{2012}.
\newblock \bibinfo{title}{A new vision on giant planet interiors: the impact of
  double diffusion convection.}
\newblock \bibinfo{journal}{Astron.~Astroph.} \bibinfo{volume}{540},
  \bibinfo{pages}{A20}.
\bibitem[{Lindal(1992)}]{Lindal92}
\bibinfo{author}{Lindal, G.}, \bibinfo{year}{1992}.
\newblock \bibinfo{title}{{The atmosphere of Neptune: an analysis of radio
  occultation data acquired with Voyager 2}}.
\newblock \bibinfo{journal}{Astronom. J.} \bibinfo{volume}{103},
  \bibinfo{pages}{967}.
\bibitem[{Lindal et~al.(1987)Lindal, Lyons, Sweetnam, Eshleman, Hinson and
  Tyler}]{Lindal+87}
\bibinfo{author}{Lindal, G.F.}, \bibinfo{author}{Lyons, J.},
  \bibinfo{author}{Sweetnam, D.}, \bibinfo{author}{Eshleman, V.},
  \bibinfo{author}{Hinson, D.}, \bibinfo{author}{Tyler, G.},
  \bibinfo{year}{1987}.
\newblock \bibinfo{title}{{The Atmosphere of Uranus: Results of Radio
  Occultation Measurements With Voyager 2}}.
\newblock \bibinfo{journal}{J.~Geophys.~Res.} \bibinfo{volume}{92},
  \bibinfo{pages}{14,987--15,001}.
\bibitem[{Lodders(2003)}]{Lodders03}
\bibinfo{author}{Lodders, K.}, \bibinfo{year}{2003}.
\newblock \bibinfo{title}{{Solar System Abundances and Condensation
  Temperatures of the Elements.}}
\newblock \bibinfo{journal}{ApJ} \bibinfo{volume}{591}, \bibinfo{pages}{1220}.
\bibitem[{Lorenzen et~al.(2009)Lorenzen, Holst and Redmer}]{Lorenzen+09}
\bibinfo{author}{Lorenzen, W.}, \bibinfo{author}{Holst, B.},
  \bibinfo{author}{Redmer, R.}, \bibinfo{year}{2009}.
\newblock \bibinfo{title}{Demixing of hydrogen and helium at megabar
  pressures}.
\newblock \bibinfo{journal}{Phys. Rev. Lett.} \bibinfo{volume}{102},
  \bibinfo{pages}{5701}.
\bibitem[{Lorenzen et~al.(2011)Lorenzen, Holst and Redmer}]{Lorenzen+11}
\bibinfo{author}{Lorenzen, W.}, \bibinfo{author}{Holst, B.},
  \bibinfo{author}{Redmer, R.}, \bibinfo{year}{2011}.
\newblock \bibinfo{title}{{Metallization in Hydrogen-Helium mixtures.}}
\newblock \bibinfo{journal}{PRB} \bibinfo{volume}{84}, \bibinfo{pages}{235109}.
\bibitem[{Marley et~al.(1995)Marley, Gomez and Podolak}]{Marley+95}
\bibinfo{author}{Marley, M.}, \bibinfo{author}{Gomez, P.},
  \bibinfo{author}{Podolak, M.}, \bibinfo{year}{1995}.
\newblock \bibinfo{title}{{Monte Carlo interior models for Uranus ad Neptune}}.
\newblock \bibinfo{journal}{J.~Geophys.~Res.} \bibinfo{volume}{100},
  \bibinfo{pages}{349--353}.
\bibitem[{Nettelmann(2009)}]{N09}
\bibinfo{author}{Nettelmann, N.}, \bibinfo{year}{2009}.
\newblock \bibinfo{title}{Matter under extreme conditions: modelling giant
  planets}.
\newblock Ph.D. thesis. Universit\"at Rostock.
\bibitem[{Nettelmann et~al.(2012)Nettelmann, Becker, Holst and Redmer}]{N+12}
\bibinfo{author}{Nettelmann, N.}, \bibinfo{author}{Becker, A.},
  \bibinfo{author}{Holst, B.}, \bibinfo{author}{Redmer, R.},
  \bibinfo{year}{2012}.
\newblock \bibinfo{title}{{Jupiter models with improved hydrogen EOS
  (H-REOS.2)}}.
\newblock \bibinfo{journal}{ApJ} \bibinfo{volume}{749}.
\bibitem[{Nettelmann et~al.(2008)Nettelmann, Holst, Kietzmann, French, Redmer
  and Blaschke}]{N+08}
\bibinfo{author}{Nettelmann, N.}, \bibinfo{author}{Holst, B.},
  \bibinfo{author}{Kietzmann, A.}, \bibinfo{author}{French, M.},
  \bibinfo{author}{Redmer, R.}, \bibinfo{author}{Blaschke, D.},
  \bibinfo{year}{2008}.
\newblock \bibinfo{title}{{Ab initio equation of state data for hydrogen,
  helium, and water and the internal structure of Jupiter}}.
\newblock \bibinfo{journal}{ApJ} \bibinfo{volume}{683}, \bibinfo{pages}{1217}.
\bibitem[{Podolak et~al.(1991)Podolak, Hubbard and Stevenson}]{Podolak+91}
\bibinfo{author}{Podolak, M.}, \bibinfo{author}{Hubbard, W.},
  \bibinfo{author}{Stevenson, D.J.}, \bibinfo{year}{1991}.
\newblock \bibinfo{title}{{Models of Uranus' Interior and Magnetic Field}}, in:
  \bibinfo{editor}{Bergstrahl, J.}, \bibinfo{editor}{Miner, E.},
  \bibinfo{editor}{Matthews, M.} (Eds.), \bibinfo{booktitle}{Uranus}.
  \bibinfo{publisher}{University of Arizona Press, Tucson},
  p.~\bibinfo{pages}{29}.
\bibitem[{Podolak et~al.(2000)Podolak, Podolak and Marley}]{Podolak+00}
\bibinfo{author}{Podolak, M.}, \bibinfo{author}{Podolak, J.I.},
  \bibinfo{author}{Marley, M.S.}, \bibinfo{year}{2000}.
\newblock \bibinfo{title}{{Further Investigation of random models of Uranus and
  Neptune}}.
\newblock \bibinfo{journal}{Planet. Space Sci.} \bibinfo{volume}{48},
  \bibinfo{pages}{143}.
\bibitem[{Podolak et~al.(1995)Podolak, Weizman and Marley}]{Podolak+95}
\bibinfo{author}{Podolak, M.}, \bibinfo{author}{Weizman, A.},
  \bibinfo{author}{Marley, M.S.}, \bibinfo{year}{1995}.
\newblock \bibinfo{title}{{Comparative models of Uranus and Neptune}}.
\newblock \bibinfo{journal}{Planet. Space Sci.} \bibinfo{volume}{43},
  \bibinfo{pages}{1517--1522}.
\bibitem[{Redmer et~al.(2011)Redmer, Mattsson, Nettelmann and
  French}]{Redmer+11}
\bibinfo{author}{Redmer, R.}, \bibinfo{author}{Mattsson, T.R.},
  \bibinfo{author}{Nettelmann, N.}, \bibinfo{author}{French, M.},
  \bibinfo{year}{2011}.
\newblock \bibinfo{title}{{The phase diagram of water and the magnetic field of
  Uranus and Neptune}}.
\newblock \bibinfo{journal}{Icarus} \bibinfo{volume}{211},
  \bibinfo{pages}{798}.
\bibitem[{Saumon et~al.(1992)Saumon, Hubbard, Chabrier and van
  Horn}]{Saumon+92}
\bibinfo{author}{Saumon, D.}, \bibinfo{author}{Hubbard, W.B.},
  \bibinfo{author}{Chabrier, G.}, \bibinfo{author}{van Horn, H.M.},
  \bibinfo{year}{1992}.
\newblock \bibinfo{title}{The role of the molecular-metallic transition of
  hydrogen in the evolution of jupiter, saturn, and brown dwarfs.}
\newblock \bibinfo{journal}{ApJ} \bibinfo{volume}{391},
  \bibinfo{pages}{827--831}.
\bibitem[{Stanley and Bloxham(2006)}]{StanBlox06}
\bibinfo{author}{Stanley, S.}, \bibinfo{author}{Bloxham, J.},
  \bibinfo{year}{2006}.
\newblock \bibinfo{title}{{Numerical dynamo models of Uranus' and Neptune's
  magnetic fields}}.
\newblock \bibinfo{journal}{Icarus} \bibinfo{volume}{184},
  \bibinfo{pages}{556--572}.
\bibitem[{Stevenson(1986)}]{Stevenson86}
\bibinfo{author}{Stevenson, D.}, \bibinfo{year}{1986}.
\newblock \bibinfo{title}{{The Uranus-Neptune dichotomy: the role of giant
  impacts}}.
\newblock \bibinfo{journal}{preprint 1986LPI...17.1011S} .
\bibitem[{Wilson and Militzer(2012a)}]{WM12a}
\bibinfo{author}{Wilson, H.F.}, \bibinfo{author}{Militzer, B.},
  \bibinfo{year}{2012}a.
\newblock \bibinfo{title}{Rock core solubility in jupiter and giant
  exoplanets}.
\newblock \bibinfo{journal}{Phys.~Rev.~Lett.} \bibinfo{volume}{108},
  \bibinfo{pages}{111101}.
\bibitem[{Wilson and Militzer(2012b)}]{WM12b}
\bibinfo{author}{Wilson, H.F.}, \bibinfo{author}{Militzer, B.},
  \bibinfo{year}{2012}b.
\newblock \bibinfo{title}{Solubility of water ice in metallic hydrogen:
  Consequences for core erosion in gas giant planets}.
\newblock \bibinfo{journal}{ApJ} \bibinfo{volume}{745}, \bibinfo{pages}{54}.

\end{thebibliography}

\clearpage
\centering

\clearpage

\clearpage

\clearpage

\clearpage

\clearpage

\clearpage
\begin{table}
\caption{\label{tab:obs}Physical constraints of Uranus and Neptune.} 
\begin{tabular}{lcccccccl}
\hline
\hline
Parameter & U data 1 & U data 2 & N data 1 & N data 2\\
\hline
$M_p$ ($\ME$)& 14.536 & 14.536 & 17.148 & 17.148\\
$\omega/2\pi$  & 17h 14m 40s$^{\rm a}$ & 16h 34m 24s$^{\rm c}$ 
               & 16h  6m 40s$^{\rm a}$ & 17h 27m 29s$^{\rm c}$\\
$T_1$ (K)      & 76(2)$^{\rm b}$ & 76(2) & 72(2)$^{\rm a}$ & 72(2)\\
$\Teff$ (K)     & 59.1(3)$^{\rm g}$ & 59.1(3) & 59.3(8)$^{\rm g}$ & 59.3(8)\\ 
\hline
$\Req'$  & 26,200$^{\rm d,e}$ & 26,200$^{\rm d,e}$ & 25,225$^{\rm d,f}$ & 25,225$^{\rm d,f}$\\
$J_2'/10^{-2}$  & 0.334129(72)$^{\rm d,e}$ & 0.334129(72)$^{\rm d,e}$   
                & 0.340843(450)$^{\rm d,f}$ & 0.340843(450)$^{\rm d,f}$ \\
$J_4'/10^{-4}$  & -0.3044(102)$^{\rm d}$ &-0.3044(102) 
                & -0.334(29)$^{\rm d,f}$ &-0.334(29)$^{\rm d,f}$\\
\hline
$\Req$ (km)    & 25,559(4)$^{\rm a}$ & 25,559(4)$^{\rm c}$ 
               & 24,766(15)$^{\rm a}$ & 24,787(4)$^{\rm c}$ \\
J$_2/10^{-2}$  & 0.351099(72)$^{\rm c}$ & 0.351099(72)$^{\rm c}$  
               & 0.35294(45)$^{\rm c}$ & 0.35294(45)$^{\rm c}$ \\
J$_4/10^{-4}$  & -0.3361(100)$^{\rm c}$ & -0.3361(100)$^{\rm c}$  
               & -0.358(29)$^{\rm c}$ & -0.358(29)$^{\rm c}$ \\
\hline
\end{tabular}

\begin{minipage}{\textwidth}
\hspace*{1cm}\\
Numbers in parenthesis are the observational error bars 
in the last digits. The gravitational moments $J_{2n}'$ are the measured ones. They refer to
a reference equatorial radius $\Req'$. The gravitational moments $J_{2n}$ 
refer to the equatorial radius at the 	1-bar pressure level, $\Req$.\\
${}^{\rm a}$\citet{Lindal92},\\ 
${}^{\rm b}$\citet{Lindal+87},\\ 
${}^{\rm c}$\citet{Helled+10},\\ 
${}^{\rm d}$http://ssd.jpl.nasa.gov,\\ 
${}^{\rm e}$\citet{BJ09},\\ 
${}^{\rm f}$\citet{Jacobson09}\\
${}^{\rm b}$after \citet{GuiGau07} 
\end{minipage}
\end{table}

\clearpage

\begin{table}
\caption{\label{tab2}Five resulting structure models.}
\begin{tabular}{lcccccl}
\hline
\hline
Parameter & U1 & U2 & N1  & N2a & N2b\\
\hline
$\Req$ (km)   & 25,559 & 25,559 & 24,773 & 24,786 & 24,786\\      
$R_{\rm mean}$ (km) & 25,388 & 25,378 & 24,622 & 24,650 & 24,650\\
$f$   & 0.0198 & 0.0210 & 0.0180 & 0.0163 & 0.0163\\
$J_2/10^{-2}$ &  0.35107 &  0.35107 & 0.3533 & 0.3531 & 0.3530\\  
$J_4/10^{-4}$ & -0.345  & -0.344 & -0.378  & -0.383 & 0.3835\\
$P_{1-2}$ (Mbar) & 0.15 & 0.15 & 0.10 & 0.30 & 1.50\\ 
$m_{1-2}$ ($M_p$) & 0.913 & 0.918 & 0.953 & 0.864 & 0.571 \\ 
$r_{1-2}$ ($R_p$) & 0.772 & 0.757 & 0.927 & 0.823 & 0.605\\ 
$Z_1$ & 0.17 & 0.08 & 0.30 & 0.60 & 0.644 \\
$Z_2$ & 0.915 & 0.944 & 0.833 & 0.805 & 1\\
$\Mcore$ ($\ME$) & 0.61 & 0.36 & 3.15 & 3.02 & 0.35\\
$\Tcore$ (K) & 6000 & 6500 & 5500 & 5000 & 5200 \\ 
$\Pcore$ (Mbar)   & 5.5 & 6.0 & 6.0 & 5.5 & 8.3 \\
$\Pcenter$ (Mbar) & 8.3 & 7.9 & 16.4 & 15.3 & 10.5 \\
$M_Z$ ($\ME$)	& 12.4 & 12.7 & 14.4 & 13.9 & 14.4 \\
I:R & 19.2 & 35.3 & 3.7 & 3.6 & 13.7\\
$\tau$ (Gyr) & 10.0 & 9.1 & 4.8 & 4.4 & 4.3 \\ 
\nI & 0.2296 & 0.2224 & 0.2405 & 0.2555 & 0.2557\\
\hline
\end{tabular}
\begin{minipage}{\textwidth}
\hspace*{1cm}\\
Uranus model U1 and Neptune model N1 are based on the input data sets No.~1, 
while models U2, N2a, and N2b are on data sets No.~2, see Table~1.
The pressure $\Pcenter$ refers to $m=0$, whereas $\Pcore$ to $m=\Mcore$.
\end{minipage}
\end{table}

\clearpage
\begin{table}
\caption{\label{tab:SOM}Example of a tabulated interior profile.} 
\begin{tabular}{ccccc}
\hline
\hline
Mass &  Pressure &  Radius & Temperature & Density\\
$(\ME)$ & (GPa) & $(\RE)$ & (K) & (g/ccm)\\
\hline
14.5322753  & 1.0000E-04  & 3.979310 &   76.0 & 4.4876E-04\\
\vdots & \vdots & \vdots & \vdots & \vdots\\
13.2697748  & 1.4999E+01  & 3.073483 & 2338.8  & 4.0541E-01\\
13.2697239  & 1.5000E+01  & 3.073465 & 2338.8  & 1.1879E+00\\
\vdots & \vdots & \vdots & \vdots & \vdots\\
0.5987030 &  5.5226E+02 &  0.700230 & 6083.5 & 4.0700E+00\\
0.5984897 & 5.5234E+02  & 0.700034  & 6083.6 & 9.0790E+00\\
\vdots & \vdots & \vdots & \vdots & \vdots\\
0.0000019 & 8.2114E+02  & 0.010000 & 6083.6 & 1.0326E+01\\
\hline
\end{tabular}

\begin{minipage}{\textwidth}
\hspace*{1cm}\\
Tables of the interior profiles of the models U1, U2, N1, and N2b are published in their 
entirety as supplemental material in the electronic edition. A portion of the table of model 
U1 is shown here for guidance of the tables' form and content. The selected rows show,
from top to bottom, the outer boundary, the outer/inner envelope transition, 
the core-mantle boundary, and the center.
\end{minipage}
\end{table}

\end{document}